\documentclass[11pt]{article}
\usepackage[utf8]{inputenc}
\usepackage[T1]{fontenc}
\usepackage[english]{babel}
\usepackage[pdftex,dvipsnames]{xcolor}
\usepackage[margin=1in]{geometry}
\usepackage{graphicx}
\usepackage{amsmath,amssymb}
\usepackage{mathtools}
\usepackage{bm}
\usepackage{booktabs}
\usepackage{multirow}
\usepackage{xargs}
\usepackage{authblk}
\usepackage{tabularx}
\usepackage[colorinlistoftodos,prependcaption,textsize=tiny]{todonotes}
\usepackage{hyperref}
\usepackage[capitalise,nameinlink,noabbrev]{cleveref}

\hypersetup{
  colorlinks=true,
  linkcolor=Plum,
  urlcolor=cyan,
  citecolor=Green
}

\newcommandx{\needref}[2][1=]{\todo[linecolor=Plum,backgroundcolor=Plum!25,bordercolor=Plum,#1]{#2}}
\newcommandx{\info}[2][1=]{\todo[linecolor=OliveGreen,backgroundcolor=OliveGreen!25,bordercolor=OliveGreen,#1]{ #2}}

\title{Commodity RF Sensing of Belowground Tuber Growth}

\author{%
Mengning Li$^{1}$, Teng Fei$^{1}$ and Wenye Wang$^{1,*}$\\
\small $^{1}$Department of Electrical and Computer Engineering, NC State University, Raleigh, NC 27606, USA\\
\small $^{*}$Correspondence: \texttt{wwang@ncsu.edu}%
}

\date{}

\begin{document}
\maketitle

\begin{abstract}

Belowground yield-forming organs of root and tuber crops are difficult to measure during growth, and management therefore relies on aboveground proxies and destructive sampling. Aboveground wireless links could provide a low-cost, non-invasive alternative, but strong attenuation and soil-dependent variability make repeatable subsurface sensing challenging. In a controlled greenhouse pot study of sweet potato, we deploy aboveground antennas in a line-of-sight-suppressed geometry and collect daily swept-frequency channel spectra together with standardized cellular link indicators, revealing consistent frequency-dependent attenuation and rippling as tubers develop. 
Here, we show that swept-frequency measurements in the 2.0--3.5 gigahertz band yield four interpretable spectral features that classify day-indexed growth stages with up to 87.5\% accuracy across two soil recipes and two moisture regimes, and that fusing cellular link-quality indicators enables 5-centimeter-grid tuber localization with up to 95.0\% accuracy, providing a proof-of-concept for subsurface crop monitoring without buried sensors, and motivating validation across cultivars and larger soil volumes.
\end{abstract}

\section*{Introduction}

\begin{figure}[t]
\centering
\includegraphics[width=\textwidth]{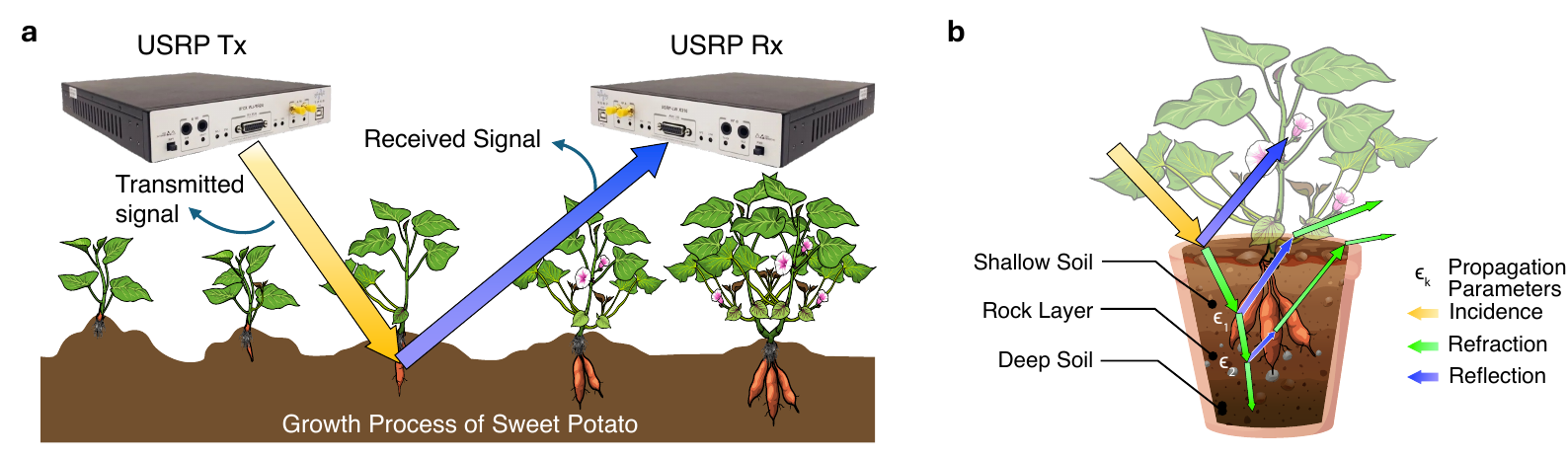}
\caption{\textbf{Overview of aboveground radio sensing of belowground tubers.} \textbf{a,} SWAMP (Subsurface Wireless Agricultural Monitoring Platform) uses aboveground radio-frequency (RF) antennas to sense tubers of sweet potato (\textit{Ipomoea batatas}) non-invasively. \textbf{b,} Conceptual propagation paths. An aluminium barrier blocks the aboveground line-of-sight (LoS) path so that received signals are dominated by energy that interacts with the pot and soil volume and tubers.}
\label{fig:system_overview}
\end{figure}

Root and tuber crops such as potato and sweet potato contribute substantially to global food and nutrition security~\cite{chandrasekara2016roots,sapakhova2023sweetpotato,ojwang2023targeting}. Yet their market-relevant biomass develops belowground and remains largely unobservable during growth, making in-season assessment difficult~\cite{li2022rootphenotyping}. As a result, crop management and breeding studies often rely on canopy-based proxies and destructive sampling, which is labor intensive, disturbs soil structure, and provides only sparse temporal snapshots of belowground development~\cite{tracy2020cropimprovement}. An overview of the sensing concept is shown in Fig.~\ref{fig:system_overview}.

Non-destructive readouts of tuber initiation, bulking, and spatial distribution could enable actionable decisions (e.g., irrigation and fertilization scheduling) and accelerate breeding by providing longitudinal belowground phenotypes without repeated excavation.
Complementary non-invasive plant sensing has been explored using minimally invasive biosensors and wireless signals for aboveground traits~\cite{bukhamsin2022biosensor,afzal2023agritera}; here we ask whether commodity radio links deployed aboveground can also provide information about subsurface storage organs.

A long-standing goal in precision agriculture and crop breeding is noninvasive subsurface phenotyping that tracks the temporal evolution of belowground biomass and spatial distribution~\cite{tracy2020cropimprovement}. However, field-ready sensing of belowground organs remains challenging because soil is lossy and heterogeneous, and its electromagnetic properties vary with moisture, texture, and temperature~\cite{topp1980em,dobson1985dielectric,or1999temperature}. As a result, environmental variability can obscure plant-induced changes.

Existing subsurface sensing approaches face practical barriers for routine deployment. Point sensors such as time-domain reflectometry (TDR) and frequency-domain reflectometry (FDR) estimate local soil dielectric permittivity and are widely used for point soil moisture sensing, but they require probe insertion and soil-specific calibration, raising concerns about durability, maintenance, and root disturbance while limiting spatial coverage~\cite{babaeian2019soilmoisture,he2023tdrreview}. Imaging approaches such as ground-penetrating radar (GPR) transmit wideband electromagnetic signals from surface antennas and infer subsurface structure from reflections, but they often require expensive instrumentation and trained operators, and performance can degrade in high-loss and heterogeneous soils and under dynamic moisture conditions~\cite{zhang2022gprreview}. Electrical resistivity tomography (ERT) can image subsurface conductivity patterns, but it similarly requires dedicated equipment and its sensitivity depends strongly on soil heterogeneity and moisture~\cite{ducut2022ertreview}.

These constraints motivate a complementary approach: leveraging commodity wireless radios deployed aboveground as a low-cost, non-invasive sensing modality for subsurface monitoring. Specifically, we ask three questions: (1) which frequency band offers the most repeatable subsurface contrast across soil texture and moisture variation; (2) can standardized cellular link-quality indicators be fused to localize and map tubers in a practical deployment geometry; and (3) can swept-frequency channel measurements support longitudinal monitoring of tuber development using interpretable channel-spectrum features.

Here, we show that aboveground wireless links in a mid-gigahertz band contain repeatable information about belowground tuber development in a controlled greenhouse pot setting. We present SWAMP (Subsurface Wireless Agricultural Monitoring Platform), which combines swept-frequency channel frequency response (CFR) spectra and standardized Long Term Evolution (LTE) link-quality indicators to enable longitudinal monitoring and coarse grid-level localization. Under a line-of-sight-blocked geometry and fixed antenna placement, SWAMP summarizes daily CFR spectra using four interpretable features and fuses five LTE indicators to produce tuber occupancy maps. Across two soil recipes and two moisture regimes in a 45~day sweet potato pot study, SWAMP achieves up to 87.5\% accuracy for day-indexed stage classification and up to 95\% accuracy for 5~cm grid localization under the tested conditions. We further relate end-of-cycle CFR feature values and temporal slopes to harvested tuber mass and volume to connect time-series signatures to plant outcome. This study is a proof-of-concept in controlled pots; extending the approach to open-field soil volumes and variable canopy and moisture conditions will require additional calibration, geometric diversity, and independent validation.
%

\begin{table}[htbp]
\centering
\renewcommand{\arraystretch}{1.2}
\setlength{\tabcolsep}{4pt}
\newcolumntype{Y}{>{\raggedright\arraybackslash}X}
\begin{tabularx}{\textwidth}{lYYYY}
\toprule
\textbf{Aspect} & \textbf{SWAMP (aboveground RF)} & \textbf{TDR probes (in-soil dielectric)} & \textbf{FDR probes (in-soil dielectric)} & \textbf{GPR (surface radar imaging)} \\
\midrule
Cost &
Low: \$1{,}000 to \$3{,}000 (two software-defined radio nodes and a host PC) &
Moderate to High: \$5{,}000 to \$10{,}000 (buried probes plus reader or logger hardware)~\cite{25AAK-tdrfdrcost} &
Moderate: about \$200 per probe plus electronics and reader or logger hardware~\cite{22Levintal-soil} &
High: \$15{,}000 to \$50{,}000 (radar unit, antennas, and proprietary software)~\cite{25APL-gprcost} \\
Setup time &
Short: 10 to 20 min (place antennas, align a repeatable geometry, and run a scripted sweep) &
Moderate: 30 to 60 min (insert probes with controlled soil contact and perform site calibration) &
Moderate: 30 to 60 min (insert probes with controlled soil contact and perform site calibration) &
Long: 1 to 3 h (survey planning, antenna coupling, calibration, and parameter tuning) \\
Performance &
Tuber monitoring (this study): up to 87.5\% for day-indexed stage classification and up to 95.0\% for 5~cm grid localization &
Point soil moisture sensing: about 95\% at the probe location after calibration~\cite{25ZZL-tdrfdracc} &
Point soil moisture sensing: sensor and soil dependent; improves with soil-specific calibration~\cite{13Vaz-calibration,15Ojo-fdr} &
Subsurface imaging: strongly soil dependent; penetration and detectability degrade in wet soils~\cite{13Guo-gprrootreview,19Zajicova-gprreview} \\
Maintenance &
Low: \$100 to \$200 per year (periodic recalibration; minimal physical wear for sensing aboveground) &
Moderate to High: \$500 to \$1{,}500 per year (recalibration; probe degradation or replacement) &
Moderate to High: \$500 to \$1{,}200 per year (recalibration; probe degradation or replacement) &
Moderate: \$300 to \$1{,}000 per year (periodic calibration; antenna and cable upkeep) \\
Spatial footprint &
Mobile scan: same hardware can be repositioned to map multiple pots or plots without in-soil sensors &
Single-point measurement per probe, so mapping requires multiple probes or repeated installations &
Single-point measurement per probe, so mapping requires multiple probes or repeated installations &
Area survey along transects or grids, with resolution and depth strongly soil and frequency dependent \\
\bottomrule
\end{tabularx}
\caption{\textbf{Deployment-oriented comparison between SWAMP and representative subsurface sensing modalities.} SWAMP uses aboveground radio-frequency (RF) measurements (swept-frequency CFR and LTE link metrics) to monitor tuber development and localize tubers on a 5~cm grid in a controlled sweet potato pot study. TDR, time-domain reflectometry; FDR, frequency-domain reflectometry; GPR, ground-penetrating radar; LTE, Long Term Evolution; CFR, channel frequency response. The performance row reports each modality's primary task rather than a single shared target metric.
Cost and maintenance values are approximate and depend on vendor configurations, scale, and local logistics.}
\label{tab:comparison}
\end{table}

\section*{Results}

We evaluate SWAMP as a proof-of-concept in a 45~day sweet potato pot study spanning two soil recipes and two moisture regimes. Conditions are denoted Soil.Moisture (Methods): SG and CB are the two soil recipes, and L1 and L2 are the two watering regimes. We organize results to support the central claim that aboveground wireless links in a mid-gigahertz band contain repeatable signatures about belowground tuber development, and that combining CFR spectra with standardized LTE indicators enables longitudinal monitoring and coarse localization within the tested setup.

Because both soil state and tuber growth jointly modulate the effective complex permittivity of the medium, the resulting propagation channel is inherently frequency dependent, motivating explicit band selection before extracting growth signatures.

We treat each pot as one independent replicate ($n=4$ pots). Daily CFR sweeps and within-position LTE samples are technical replicates used to reduce measurement noise (Methods).

Table~\ref{tab:comparison} summarizes deployment-oriented tradeoffs between SWAMP and representative subsurface sensing modalities. The goal is to contrast invasiveness, setup overhead, and spatial footprint for the primary tasks each modality is typically used for in practice.

\begin{figure}[tbp]
\centering
\includegraphics[width=.85\textwidth]{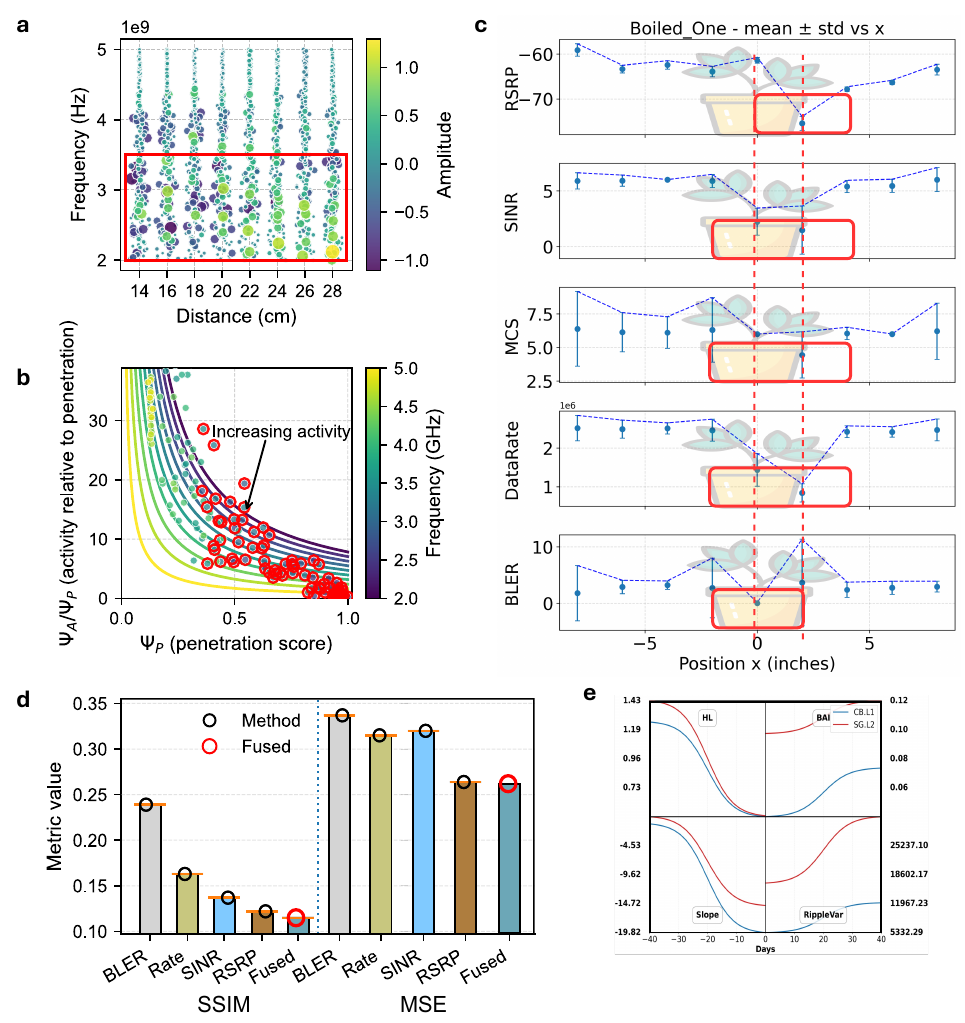}
\caption{\textbf{Experimental validation of band selection, sensing metrics, localization, and growth monitoring.}
\textbf{a,} Channel frequency response (CFR) samples over antenna separation, where the most stable contrast between the soil-only case and the soil-plus-tuber case concentrates in 2.0 to 3.5~GHz (red box; color indicates normalized amplitude).
\textbf{b,} Penetration score $\Psi_P$ versus relative activity $\Psi_A/\Psi_P$ derived from the attenuation model, illustrating the penetration and sensitivity tradeoff across frequency (color indicates frequency).
\textbf{c,} Long Term Evolution (LTE) link quality indicators measured along the lateral scan axis $x$, reported as mean with standard deviation.
(Error bars show s.d.; $n=100$ samples per position from a 20~s window at 5~Hz unless stated otherwise.)
Reference signal received power (RSRP), signal to interference plus noise ratio (SINR), modulation and coding scheme (MCS), data rate, and block error rate (BLER) are shown. Vertical dashed lines mark the ground truth tuber span and red boxes highlight the region where single metrics degrade relative to the fused estimate.
\textbf{d,} Localization agreement with ground truth quantified by structural similarity index (SSIM) and mean squared error (MSE), where linear fusion improves SSIM and reduces MSE compared with any single LTE metric.
\textbf{e,} Representative CFR derived features over the 45~day growth cycle for two soil and moisture conditions, showing consistent temporal evolution in H/L, BAI, Slope, and RippleVar.
(Each curve shows one pot; $n=45$ days per pot.)}
\label{fig:results_overview}
\end{figure}

\subsection*{A mid gigahertz band maximizes subsurface sensitivity}

An aboveground radio link probing soil faces an immediate tradeoff: lower frequencies penetrate deeper but require larger antennas and provide weaker sensitivity to small-scale dielectric structures. Higher frequencies offer stronger sensitivity to fine structure but attenuate rapidly in moist and conductive media. To stay compatible with compact aboveground antennas and commodity RF front ends, we focus on the mid gigahertz range and empirically determine where the subsurface contrast is strongest.
We quantify subsurface sensitivity using the channel frequency response (CFR), the complex transfer function $H(f)$ of the propagation channel as a function of frequency. Essentially, CFR perturbations reflect a superposition of plant induced dielectric changes and soil driven variability (e.g., moisture and conductivity), so we seek a band where the tuber induced contrast remains repeatable across conditions. Its magnitude reflects frequency dependent attenuation from soil loss and biomass absorption. Its phase reflects frequency dependent delay and scattering due to multipath created by interfaces such as air to soil boundaries and tuber induced discontinuities.

Across conditions, the most repeatable contrast concentrates in a contiguous band from 2.0 to 3.5~GHz, with a pronounced absorption trough centered near 2.68~GHz (Fig.~\ref{fig:results_overview}a). Above about 3.5~GHz, attenuation trends largely plateau, indicating that additional frequency increase contributes limited incremental information about tuber induced change while increasing sensitivity to loss. We therefore fix 2.0 to 3.5~GHz as the working band for the remainder of the paper. This working band balances penetration and sensitivity: it is high enough to capture wavelength scale scattering from emerging tuber structure, yet low enough to retain usable penetration under moisture varying soil.

To connect this empirical choice to an interpretable mechanism, we evaluate a penetration sensitivity tradeoff predicted by an attenuation model of the air to pot to soil to tuber medium. Specifically, Fig.\ref{fig:results_overview}b plots the penetration score $\Psi_P$ against the relative activity $\Psi_A/\Psi_P$ across frequency. (Definitions of $\Psi_P$ and $\Psi_A$ and the multilayer model parameters are provided in Methods.) Lower frequencies yield higher $\Psi_P$ but lower relative activity, reflecting improved penetration with weaker sensitivity to localized dielectric discontinuities. Higher frequencies increase relative activity but incur rapidly reduced penetration, especially under wetter and more conductive conditions. The selected 2.0 to 3.5~GHz band lies near the transition region where penetration remains sufficient for the pot geometry while the predicted activity increases, consistent with the stable contrast observed in the CFR samples in Fig.~\ref{fig:results_overview}a, in our pot study.

\subsection*{Cellular signal fusion enables tuber localization}

\begin{figure}[tbp]
    \centering
    \includegraphics[width=0.98\linewidth]{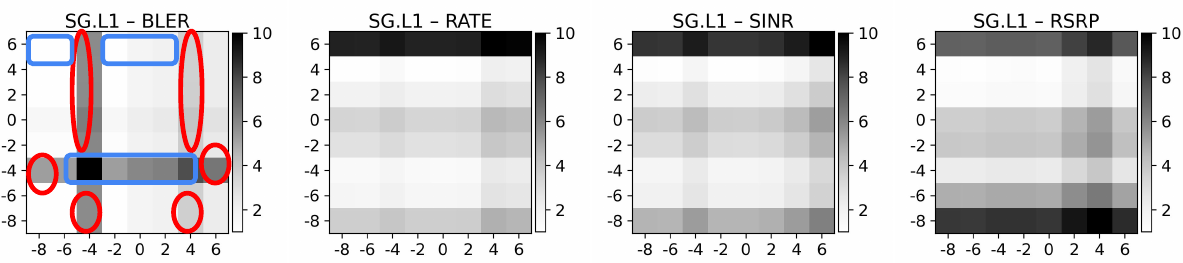}
    \caption{A single metric cannot fully represent the ground-truth heatmap. Using the BLER generated heatmap as an example, the blue rectangles indicate regions where the reconstructed heatmap matches the ground truth, while the red rectangles indicate regions that do not. This mismatch motivates the need for a fusion metric to generate heatmaps that better align with the ground truth.}
    \label{fig:singlemetricheatmap}
\end{figure}

Band limited CFR measurements support longitudinal monitoring at a fixed geometry. While the underlying mechanism remains dielectric modulation of propagation, localization requires spatial diversity because each link quality indicator provides only a partial projection of the same channel perturbation under a given geometry. Coarse localization instead requires geometric diversity. We therefore leverage Long Term Evolution (LTE), a widely deployed cellular standard that estimates link quality for link adaptation and exposes standardized indicators on commodity radios.

In our controlled private LTE setup, we raster scan a planar grid around each pot and record five LTE indicators at each grid cell over repeated short windows. Reference signal received power (RSRP) captures received signal strength on LTE reference symbols. Signal to interference plus noise ratio (SINR) captures link quality. The modulation and coding scheme index (MCS) captures the chosen spectral efficiency based on channel conditions. Throughput reports realized data rate after adaptation. Block error rate (BLER) captures decoding failures.

Empirically, no single LTE metric consistently reconstructs tuber location. Along the lateral scan axis $x$, individual indicators exhibit local degradations over the ground truth tuber span and can disagree in where and how strongly the link quality changes (Fig.~\ref{fig:results_overview}c). For example, under SG.L1, MCS achieves 12.5\% accuracy within a 5.1~cm (2~inch) tolerance margin and 15.6\% within a 10.2~cm (4~inch) margin, while RSRP reaches 37.5\% and 50\% under the same margins. Across all conditions, individual metric accuracy spans 12.5\% to 61\%, indicating that each indicator captures only a partial projection of the underlying propagation changes.

To obtain more reliable localization in this controlled geometry, we fuse the five LTE metrics into a single score map using a simple linear model with nonnegative weights constrained to sum to one Since a single metric has been shown to be insufficient to accurately reconstruct the heatmap relative to the ground truth(Fig.~\ref{fig:singlemetricheatmap}). Aggregated over the dataset, fusion improves agreement with ground truth, increasing structural similarity index (SSIM) while reducing mean squared error (MSE) relative to any single LTE metric (Fig.~\ref{fig:results_overview}d). Validated on an independent dataset collected with the same protocol (Methods), the fused map achieves up to 95\% localization accuracy, with the lowest performing condition still reaching 87.5\%. Fusion also reduces continuous map error, with MSE dropping to 0.115 compared with 0.122 to 0.239 for individual metrics.
Because occupancy maps are sparse (many grid cells are empty), we report SSIM and MSE alongside accuracy and use tolerance-dilated accuracy to reduce sensitivity to small annotation shifts (Methods).

\subsection*{CFR features enable longitudinal growth monitoring}

Finally, we evaluate SWAMP for longitudinal growth monitoring under a fixed aboveground geometry. With the 2.0 to 3.5~GHz working band fixed, we use daily swept frequency CFR measurements to track subsurface development over time. As tuber biomass develops, the CFR magnitude exhibits increased frequency selective loss and stronger fine scale rippling, consistent with higher absorption and additional scattering interfaces within the soil volume.

To summarize temporal change, we compute four interpretable CFR features over 2.0 to 3.5~GHz (Methods). Broadband attenuation integral (BAI) captures aggregate loss across the band. The high to low band ratio (H/L) captures differential attenuation of higher frequencies. Slope captures the fitted frequency dependence of log magnitude. Ripple Variance~\cite{13AL-ripple,25NOH-ripple} (RippleVar) captures fine scale spectral ripples associated with scattering.

Across the 45~day study, these CFR derived features evolve reproducibly within each pot despite changes in soil recipe and moisture regime (Fig.\ref{fig:results_overview}e). The direction of change is consistent with the physical meaning of each feature. As tubers grow, higher effective water content and permittivity increase absorption, reducing $\left|H(f)\right|$ and increasing BAI. Because tubers attenuate higher frequency components more strongly, H/L decreases as biomass increases. High frequency attenuation increases more substantially in wetter media, making the fitted Slope more negative as growth progresses. As tubers reach sizes comparable to the RF wavelength (approximately 3 to 15~cm over 2.0 to 3.5~GHz), internal reflections and additional scattering paths introduce spectral ripples, which are captured by RippleVar, across pots and conditions.

Because the day indexed stage labels used for monitoring are a proxy for development (Methods), we additionally relate these CFR signatures to harvested biomass outcomes at the end of the 45~day period. For each pot, we compute the end of cycle feature value at day~45. We also compute a temporal trend estimated from the feature trajectory over days. We then correlate these descriptors with harvested tuber mass and volume measured at day~45. Across the four pots, the associations are directionally consistent with the feature interpretations, but the small number of biologically independent replicates makes the effect estimates uncertain. We therefore report correlation coefficients with bootstrap confidence intervals and treat this analysis as exploratory support for outcome relevance rather than a definitive calibration (Methods).

As an operational proxy for in season monitoring under a fixed geometry, we also formulate a supervised day indexed stage classification task using the same four CFR features. Using multinomial logistic regression and leave one pot out evaluation (Methods), SWAMP classifies early, middle, and late day windows with up to 87.5\% accuracy across the tested conditions.

\section*{Discussion}

\begin{figure}[tbp]
\centering
\includegraphics[width=.98\textwidth]{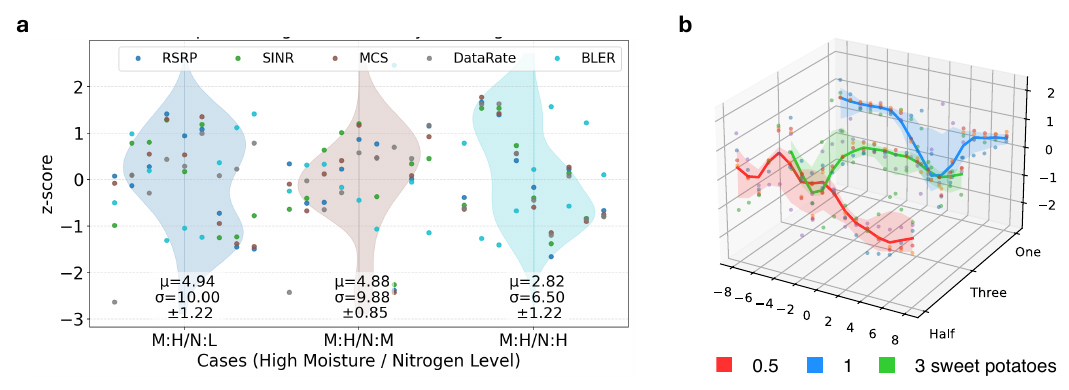}
\caption{\textbf{LTE condition sensitivity and tuber counting signatures.}
\textbf{a,} Distributions of standardized LTE link indicators under three high moisture cases with increasing nitrogen level (N:L, N:M, N:H). Violin plots show the empirical density and overlaid points show individual measurements for RSRP, SINR, MCS, data rate, and BLER after z-score standardization, illustrating condition dependent shifts and within case variability.
\textbf{b,} Grid aggregated LTE signatures for tuber counting across lateral position $x$ for three occupancy settings (0.5, 1, and 3 sweet potatoes). Curves show mean trends with shaded variability over repeated scans, demonstrating separable responses for sparse versus dense occupancy while the intermediate case is less consistently separated.}
\label{fig:discussion}
\end{figure}

RF based subsurface sensing leverages the fact that the measured wireless channel is a deterministic function of the electromagnetic propagation medium, which in this setting is governed by the coupled dielectric properties of soil and the developing sweet potato tuber. As the tuber grows, its increasing volume and water rich tissue introduce a localized perturbation in the effective complex permittivity within the soil bed, thereby modifying wave attenuation, scattering, and multipath interference. These changes manifest as systematic variations in channel observables such as frequency dependent channel frequency response (CFR), received signal strength, and other link-quality indicators derived from the baseband channel. However, the same observables are also strongly modulated by environmental factors, such as most notably soil moisture and nutrient composition (e.g., nitrogen), which shift the dielectric constant and conductivity and thus reshape the propagation channel even in the absence of growth. In addition, the sensitivity of RF measurements to these mechanisms is inherently frequency dependent due to penetration depth, wavelength scale scattering, and dispersion effects. Therefore, accurately inferring tuber size from RF signatures requires a sensing formulation that jointly models frequency selection and disentangles plant induced channel perturbations from confounding soil-driven variability, motivating multi-metric and multi-condition RF sensing for robust underground sensing.

In addition to longitudinal monitoring and localization, we also explored whether soil management factors and tuber condition leave learnable signatures in aboveground radio measurements. We varied soil nutrient level through nitrogen amendment under controlled moisture settings, and we evaluated tuber physical state as a proxy for belowground health by comparing fresh tubers with a boiled tuber. For these analyses, we standardize each wireless metric using z-score normalization so that shifts in distribution reflect relative changes rather than raw scale differences across indicators. As shown in Fig.~\ref{fig:discussion}a, the distributions of z-scored link quality metrics shift across nitrogen levels under high moisture, indicating that nutrient driven changes in the soil medium can measurably alter propagation and therefore must be treated as a confounder when interpreting growth related trends. This observation motivates condition aware modeling and reinforces the role of normalization and fusion when comparing across management regimes.


We also analyzed tuber amount and tuber physical state using standardized \textbf{LTE-derived} metrics across configurations with a half sweet potato, three sweet potatoes, and one boiled sweet potato. As illustrated in Fig.~\ref{fig:discussion}b, the boiled sweet potato produces the lowest z-score values, consistent with altered tissue composition and water distribution that reduce dielectric contrast and scattering relative to fresh tubers. In these controlled burials, the \textbf{three-sweet-potato} configuration produces the strongest z-score shifts, while the \textbf{half-sweet-potato} configuration is intermediate. We interpret this ordering as evidence that the sensing stream is sensitive to both tuber quantity and tuber state; however, because responses can also depend on placement depth, coupling to pot boundaries, and local moisture heterogeneity, we do not assume a strictly monotonic mapping from ``more biomass'' to ``larger metric shift'' without additional geometric diversity and controlled placement studies. Building on this separability, we treat tuber number as a classification problem, where aggregated standardized metrics are used to predict the occupancy class rather than regressing an exact count.

This work provides a proof of concept that aboveground radio measurements can recover temporal cues and coarse spatial structure of belowground tuber development without embedding electronics in soil. In a controlled greenhouse pot environment with a line of sight suppressed geometry, SWAMP combines two complementary views of the same subsurface medium. Swept frequency CFR spectra are summarized into interpretable features that evolve reproducibly over days and support longitudinal monitoring across conditions. In parallel, standardized LTE link indicators provide a second sensing stream that can be fused into coarse occupancy maps on a centimeter scale grid for tuber localization and counting in practice.

An important practical question is whether these wireless indicators remain usable under agronomic variability, because soil water and nutrient conditions can shift propagation in ways that mimic biomass driven change. Under high moisture, we observe that standardized LTE indicators exhibit systematic distribution shifts across nitrogen levels with substantial within case spread (Fig.~\ref{fig:discussion}a). This result implies that single indicators should be treated as condition dependent proxies and that fusion is necessary to support comparisons across management regimes.

Beyond sensitivity to environmental conditions, we evaluate whether the LTE stream contains information about tuber number at pot scale. When aggregating scans across the grid, the reconstructed signatures exhibit separable trends between single tuber and three tuber configurations, while the intermediate condition is less consistently separated when spatial responses flatten due to coupling and boundary effects (Fig.~\ref{fig:discussion}b). These results support coarse counting, but they also indicate that finer count resolution will likely require increased geometric diversity and additional priors that constrain the spatial map.

At the same time, the current evidence is bounded by the experimental design. The study is limited to a small number of pots in a greenhouse and a narrow set of soil and watering conditions. Repeated measurements provide high temporal resolution, but they do not substitute for additional biologically independent replicates or growth cycles, so correlations to harvest outcomes should be interpreted as exploratory. In addition, the measurement geometry intentionally suppresses aboveground multipath to emphasize soil interacting energy, so open field settings will introduce confounders from canopy growth and variable antenna coupling that are not yet captured here.

Taken together, these observations clarify what SWAMP does and does not yet solve. The platform demonstrates repeatable information content in a mid gigahertz band for growth monitoring and shows that LTE metric fusion can support coarse localization and counting in a controlled pot scale setup. It does not yet establish field scale generalization or calibration procedures for heterogeneous agricultural soils. A practical translation will likely hinge on explicit handling of soil state variation, improved pose robustness through more diverse measurement geometry, and validation across cultivars and management regimes, ideally with mid season destructive sampling to anchor time series signatures to biomass trajectories rather than day index proxies.

\section*{Methods}

\subsection*{Experimental design and plant growth}

\begin{figure}[tbp]
\centering
\includegraphics[width=.98\textwidth]{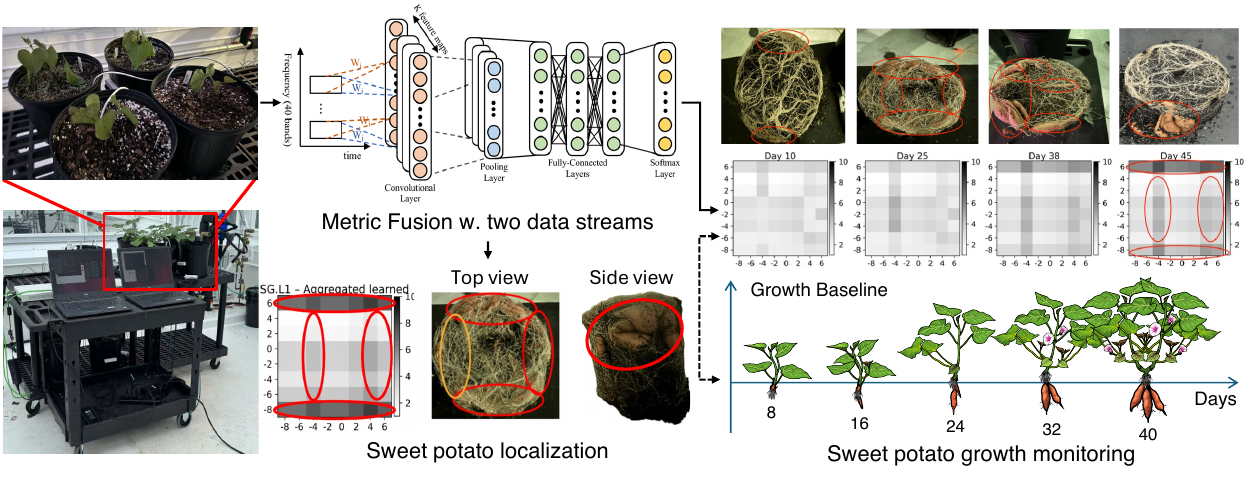}
\caption{\textbf{System overview of SWAMP for sweet potato monitoring with two wireless data streams.}
A cart based measurement platform performs repeated scans of potted plants in a greenhouse.
Swept frequency channel frequency response (CFR) is collected using a software defined radio (SDR) link and converted into CFR derived features for longitudinal growth stage monitoring.
In parallel, cellular Long Term Evolution (LTE) link quality metrics are sampled over a planar grid and fused with a constrained linear model to produce a spatial probability map for tuber counting and localization.
Representative excavations and reconstructed maps illustrate the growth baseline across days.}
\label{fig:method_overview}
\end{figure}

Sweet potato cuttings were cultivated in four identical 3~gallon pots (height 10~in, diameter 8~in) with one cutting per pot.
Day~1 is defined as the planting day and is also the first day of radio measurement.
%
We evaluated two soil recipes, Sunshine Propagation Mix is denoted SG and JollyGardener HFCB Mix is denoted CB~\cite{ncsu_phytotron_substrates}.
According to the facility description, SG is a peat and perlite substrate with dolomite lime and a long lasting wetting agent, and CB contains 25\% peat moss, 39\% pine bark, 8\% perlite, and 28\% HydraFiber or HFEZ~\cite{ncsu_phytotron_substrates}.

Each soil recipe was studied under two moisture regimes denoted L1 and L2, yielding four conditions SG.L1, SG.L2, CB.L1, and CB.L2.
Each pot had an independent water supply. Irrigation was performed twice per day at 08:00 and 17:00, with each session lasting 5~min. For the L1 moisture condition, water was sprayed uniformly over 360$^\circ$ around the pot. For L2, water was applied over a 90$^\circ$ sector, starting from the edge of the pot and directed toward the sweet potato stem to create a consistent asymmetric moisture distribution.
Water was applied using individual hoses terminated with adjustable spray nozzles so that the angular coverage could be reproduced between sessions.

Plants were grown in the NC State Phytotron temperature controlled greenhouses.
The greenhouses are unshaded and the radiant flux density inside is approximately 80\% to 88\% of outside natural light conditions, depending on solar azimuth~\cite{ncsu_phytotron_manual}.
At night, the facility can impose long day conditions by interrupting the dark period from 23:00 to 02:00 using low intensity incandescent lamps at approximately 11 to 12~\(\mu\)mol~s\(^{-1}\)~m\(^{-2}\)~\cite{ncsu_phytotron_manual}.
Air temperature was maintained at 27.3~\(^\circ\)C and relative humidity was maintained at 64\% (nominal constant throughout the study).
All radio measurements were taken once per day at 19:00 over a 45~day period.

\textbf{Harvest outcomes.} At harvest on day~45, tubers were excavated and recorded for both spatial ground truth and biomass outcomes. Tubers were photographed in situ before removal, and tuber centroids were annotated in a pot fixed coordinate system and rasterized into a binary occupancy map on a 5~cm grid for localization ground truth.
Tubers were rinsed, blotted dry, and weighed on a digital scale. Volumes were measured by water displacement in a graduated cylinder.

We define the pot fixed coordinate system in the horizontal plane with the origin at the pot centre, the $x$ axis aligned with the transmitter to receiver direction, and the $y$ axis orthogonal to $x$ in the plane.
Centroid coordinates were converted from image pixels to centimeters using a scale reference in each photograph and then quantized to 5~cm grid indices by integer division.
A grid cell was labeled occupied if at least one tuber centroid fell within the cell.

\subsection*{Localization with cellular metrics}

SWAMP localizes tubers by raster scanning a planar grid around each pot and recording standardized LTE link indicators at each grid cell. We then fuse the normalized indicators into a single score map using a constrained linear model.

\textbf{LTE acquisition.} LTE acquisition is implemented using a modified version of srsRAN (Release v22.04) configured as a private LTE system~\cite{srsran}. A transmitter configured as an evolved NodeB (eNodeB) communicates with a receiver configured as user equipment (UE). Both the eNodeB and UE run on Dell Precision 7750 laptops connected to USRP X310 devices via RJ 45 Ethernet interfaces. Each laptop is equipped with an Intel Core i7 10875H processor (8 cores, up to 5.1~GHz), a 1~TB SSD, and 16~GB RAM.
The LTE base station and UE are configured with a centre frequency of 2680~MHz. The number of physical resource blocks is set to $n_{\mathrm{prb}}=50$, corresponding to a 10~MHz system bandwidth. The LTE baseband sampling rate is 1.92~MS/s. The LO frequency is set equal to the configured centre frequency.
%
At each receiver position, we maintain the LTE link for 20~s and record link quality indicators at 5~Hz, yielding 100 samples per position. We record reference signal received power. We record signal to interference plus noise ratio. We record modulation and coding scheme. We record block error rate. We record throughput.

\textbf{Cellular grid mapping and metric fusion.} For localization experiments, the UE antenna is repositioned over a planar measurement grid around each pot with 5~cm spacing between adjacent positions.
At each grid cell, we maintain the LTE link for 20~s and record the five indicators at 5~Hz. We discard the first 1~s to reduce transient adaptation effects and aggregate the remaining samples per metric using the median.
Each metric heatmap is normalized per pot by subtracting the across grid mean and dividing by the across grid standard deviation.

We fuse the five normalized LTE heatmaps into a single score map using a constrained linear model. Let $g$ index grid cells, $x_j(g)$ denote metric $j$ at cell $g$, and $y(g)\in\{0,1\}$ denote the ground truth occupancy label. We predict a score
\begin{equation}
\widehat{y}(g) = \sum_{j=1}^{5} w_j\, x_j(g) .
\end{equation}
Weights are constrained to be nonnegative and to sum to one. We parameterize weights using a softmax transform,
\begin{equation}
w_j = \frac{\exp(u_j)}{\sum_{k=1}^{5}\exp(u_k)} .
\end{equation}
We learn $u$ by minimizing mean squared error,
\begin{equation}
\mathcal{L}(u)
=
\frac{1}{|\mathcal{G}|}
\sum_{g\in\mathcal{G}}
\left(
\sum_{j=1}^{5} w_j(u)\, x_j(g) - y(g)
\right)^2 ,
\end{equation}
where $\mathcal{G}$ is the set of training grid cells. Optimization is performed using BFGS~\cite{nocedal2006}.
For visualization and thresholding, $\widehat{y}(g)$ is rescaled to $[0,1]$ within each pot.

\subsection*{Growth monitoring with CFR}
Channel frequency response (CFR) is the complex transfer function $H(f)$ of the propagation channel as a function of frequency. The CFR magnitude captures frequency-dependent attenuation due to soil loss and biomass absorption, while the CFR phase captures frequency-dependent delay and scattering from multipath created by interfaces such as air--pot--soil boundaries and tuber-induced discontinuities. We repeat this definition here (in addition to Results) to make the signal representation self-contained in Methods.
SWAMP monitors tuber development over time using daily swept frequency CFR measurements collected at a fixed aboveground geometry. CFR features are computed within the selected working band and used for longitudinal tracking, stage classification, and association with harvest outcomes.

\textbf{CFR acquisition.} CFR acquisition is implemented using a pair of software defined radios built from Universal Software Radio Peripheral (USRP) X310 devices controlled through GNU Radio~\cite{gnuradio,usrpx310}. To reduce hardware dependencies, we synchronize the transmitter and receiver using a lightweight Transmission Control Protocol (TCP) coordination protocol over Ethernet rather than external clock distribution.
Each USRP X310 connects to its host computer via 1~Gb Ethernet. At each measurement session, the transmitter generates a unit amplitude sinusoidal tone and steps the tone frequency from 2.0~GHz to 5.0~GHz in 40~MHz increments. At each frequency step, the system waits 0.3~s for front end settling and then records 0.2~s of complex baseband samples before hopping to the next frequency. The complex baseband sampling rate is 500~kS/s. The local oscillator frequency is set equal to the instantaneous centre frequency at each step. The USRP transmit gain is set to 60~dB (device gain setting). We use NI VERT2450 antennas on all RF chains, and we connect both RF chains of the USRP X310 for consistency checks across chains.

To ensure repeatable geometry across days without disturbing soil, the pots were placed on one wheeled cart and the SDR radios and antennas were mounted on a second wheeled cart. The USRPs were positioned with their antennas facing each other, with pot bases aligned at the same height. The antennas were mounted 8~in above the ground (pots are 10~in tall), and the separation between the two antennas was 10~in. A 10~in high aluminium barrier was placed between the antennas to block the line of sight path aboveground, forcing the dominant energy to interact with the pot and soil medium before reaching the receiver.
Assuming the pot is centred between antennas, the distance from each antenna phase centre to the nearest pot wall along boresight is approximately 1~in based on the 10~in antenna separation and 8~in pot diameter.

The sweep bandwidth motivates the sensing design from a physics perspective. A wider bandwidth improves depth resolution, while higher frequencies reduce penetration in lossy soil. Using the real part of relative permittivity $\epsilon_r'$ as an effective wave speed factor, a common depth resolution approximation is given by
\begin{equation}
d_{\mathrm{res}} \approx \frac{c}{2 B \sqrt{\epsilon_r'}} ,
\end{equation}
where $c$ is the speed of light and $B$ is swept bandwidth.
Penetration can be summarized through the attenuation constant $\alpha(f)$ and power penetration depth $\delta_p(f)$:
\begin{equation}
\delta_p(f) \approx \frac{1}{\alpha(f)} ,
\quad
\alpha(f) = \mathrm{Re}\{\gamma(f)\} ,
\quad
\gamma(f) = j 2\pi f \sqrt{\mu_0 \epsilon_0 \tilde{\epsilon}_r(f)} ,
\end{equation}
where $\tilde{\epsilon}_r(f)$ is the complex relative permittivity, $\mu_0$ is vacuum permeability, and $\epsilon_0$ is vacuum permittivity.

\textbf{Physics guided multilayer model.} To interpret band selection and generate testable predictions, we model the measurement geometry as a multilayer electromagnetic medium comprising air, pot wall, soil, and a localized high water content inclusion representing a tuber.

\textbf{CFR preprocessing and growth features.} For a stepped frequency tone at frequency $f$, the received complex baseband can be approximated as
\begin{equation}
r_f(t) = H(f)\, e^{j 2\pi \Delta f\, t} + n(t) ,
\end{equation}
where $H(f)$ is the channel frequency response at frequency $f$, $\Delta f$ is residual carrier frequency offset, and $n(t)$ is noise plus interference. We estimate and remove $\Delta f$ by fitting a linear phase ramp over time and compensating it before estimating $H(f)$. After compensation, we compute the channel estimate as the time average of the complex samples within the recording window,
\begin{equation}
\widehat{H}(f) = \frac{1}{T}\int_{0}^{T} r_f(t)\, e^{-j 2\pi \widehat{\Delta f}\, t}\, dt .
\end{equation}
We represent CFR in polar form,
\begin{equation}
\widehat{H}(f) = \left| \widehat{H}(f) \right| e^{j \widehat{\phi}(f)} ,
\quad
A_{\mathrm{dB}}(f) = 20 \log_{10}\left| \widehat{H}(f) \right| ,
\quad
\widehat{\phi}(f) = \mathrm{unwrap}\bigl(\angle \widehat{H}(f)\bigr) .
\end{equation}

CFR structure can be interpreted through a multipath expansion,
\begin{equation}
H(f) = \sum_{k=1}^{K} a_k\, e^{-j 2\pi f \tau_k} ,
\end{equation}
where $a_k$ and $\tau_k$ are the complex attenuation and delay of the $k$th path.

To reduce static front end ripple, we record an air reference sweep with the same antenna separation and flatten each pot sweep by complex division. For longitudinal comparisons within a pot, each day is normalized relative to day~1 by complex division after air reference flattening.
Concretely, letting $\widehat{H}_{\mathrm{pot},d}(f)$ be the estimated CFR on day $d$ and $\widehat{H}_{\mathrm{air}}(f)$ the air reference, we compute $\widehat{H}^{(1)}_d(f)=\widehat{H}_{\mathrm{pot},d}(f)/\widehat{H}_{\mathrm{air}}(f)$ and then $\widehat{H}^{\mathrm{norm}}_d(f)=\widehat{H}^{(1)}_d(f)/\widehat{H}^{(1)}_1(f)$.

Unless stated otherwise, growth features are computed over the selected working band from $f_1 = 2.0$~GHz to $f_2 = 3.5$~GHz at discrete frequencies $\{f_k\}_{k=1}^{N}$ with step $\Delta f$.

We compute four CFR features. Broadband attenuation integral (BAI) quantifies total attenuation area relative to the maximum magnitude in band,
\begin{equation}
\mathrm{BAI}
=
\frac{1}{f_2 - f_1}
\int_{f_1}^{f_2}
\left(
A_{\max} - A_{\mathrm{dB}}(f)
\right)\, df ,
\quad
A_{\max} = \max_{f\in[f_1,f_2]} A_{\mathrm{dB}}(f) ,
\end{equation}
and we approximate the integral by a discrete sum.

High to low band ratio (HL) splits the working band into a low band $[2.0, 2.75]$~GHz and a high band $[2.75, 3.5]$~GHz and computes the ratio of integrated linear magnitudes,
\begin{equation}
\mathrm{HL}
=
\frac{\int_{2.75}^{3.5} \left| \widehat{H}(f) \right|\, df}
{\int_{2.0}^{2.75} \left| \widehat{H}(f) \right|\, df} .
\end{equation}

Spectral slope fits a line to $A_{\mathrm{dB}}(f)$ over $[f_1,f_2]$ using least squares. The slope coefficient is
\begin{equation}
\mathrm{Slope}
=
\frac{\sum_{k=1}^{N} (f_k - \overline{f})\left(A_{\mathrm{dB}}(f_k) - \overline{A}_{\mathrm{dB}}\right)}
{\sum_{k=1}^{N} (f_k - \overline{f})^2} ,
\end{equation}
where $\overline{f}$ and $\overline{A}_{\mathrm{dB}}$ are means over the $N$ samples.

Ripple variance constructs a smoothed baseline $\widetilde{A}(f)$ from the linear magnitude spectrum using median filtering and measures normalized ripple strength,
\begin{equation}
\mathrm{RippleVar}
=
\mathrm{std}\!\left(
\frac{\left| \widehat{H}(f) \right|}{\widetilde{A}(f)}
\right)_{f\in[f_1,f_2]} .
\end{equation}

For interpreting soil moisture effects, we map phase slope to apparent permittivity and then to volumetric water content. A standard approximation links phase slope to group delay
\begin{equation}
\tau_g(f) = -\frac{1}{2\pi}\frac{d\widehat{\phi}(f)}{df} .
\end{equation}
An effective phase velocity estimate can be expressed as $v \approx d_{\mathrm{eff}}/\tau_g$ for an effective path length $d_{\mathrm{eff}}$.
We then compute apparent permittivity
\begin{equation}
\epsilon_p = \left(\frac{c}{v}\right)^2 ,
\end{equation}
and map $\epsilon_p$ to volumetric water content $\theta_v$ using the Topp relation~\cite{topp1980em}
\begin{equation}
\theta_v
=
4.3\times 10^{-6}\epsilon_p^3
-
5.5\times 10^{-4}\epsilon_p^2
+
2.92\times 10^{-2}\epsilon_p
-
5.3\times 10^{-2} .
\end{equation}
Because greenhouse temperature was controlled at 27.3~$^\circ$C, we treat temperature as constant during the experiment.

\textbf{Growth stage labelling and classification.} We formulate longitudinal monitoring as a supervised day indexed stage classification task.
Because belowground tuber development is not directly observable non destructively during the 45~day period, we assign stage labels using the day index after planting as a practical proxy.
We partition the observation window into three stages with equal duration.
Days 1 to 15 are labelled as the early stage.
Days 16 to 30 are labelled as the middle stage.
Days 31 to 45 are labelled as the late stage.
We emphasize that these labels are \emph{time bins} rather than phenologically validated stages; future work will incorporate staged destructive sampling or imaging to label tuber initiation and bulking directly.

Each day contributes one CFR sweep and one feature vector computed from the working band.
Broadband attenuation integral captures overall attenuation across the band.
High to low band ratio captures differential attenuation between the upper and lower portions of the band.
Spectral slope summarises the frequency trend of attenuation.
Ripple variance quantifies fine scale spectral structure linked to additional scattering interfaces.

Before training, each feature is standardised using the mean and standard deviation computed on the training set only.
We train a multinomial logistic regression classifier with L2 regularisation.
The regularisation strength is selected by five fold cross validation within the training set.
To evaluate generalisation across soil and watering conditions, we use leave one pot out cross validation.
In each fold, three pots are used for training and the held out pot is used for testing.
We report classification accuracy averaged across the four folds.
Unless otherwise stated, the dataset includes one sweep per day for each pot over 45 days, which yields 180 samples in total.

\textbf{CFR feature association with harvested biomass.} To relate time series CFR signatures to harvest outcomes, we compute per pot two descriptors for each feature. We compute the end of cycle value at day~45. We compute a temporal slope estimated by least squares regression of feature versus day index across the 45~day window. We then compute Pearson and Spearman correlation coefficients between these descriptors and harvested tuber mass and volume across pots. Because the number of pots is small ($n=4$ biologically independent replicates), we report bootstrap confidence intervals (10{,}000 resamples with replacement at the pot level) and treat $p$ values as descriptive.
Slopes are computed over the full 45~day window using ordinary least squares.

\subsection*{Soil condition monitoring with cellular metrics}

To characterize how cellular indicators respond to changes in subsurface composition and environment beyond the 45~day growth study, we conduct controlled perturbation experiments using the LTE setup and indicators described above. We keep the antenna geometry and acquisition settings fixed. We then vary the underground target and the soil environment in isolation. We vary tuber amount by placing one half tuber in soil and by placing three fresh tubers in soil. We vary tuber state by placing a boiled tuber in soil. We further conduct experiments under three controlled soil moisture levels and three nitrogen concentrations to evaluate sensitivity to environmental conditions.

For each condition, we record the same five LTE indicators used for localization and aggregate samples within each time window using the same protocol. We pool the aggregated values across repeated trials and, when applicable, across grid cells to obtain a distribution per metric under each condition. We visualize these distributions as probability density functions in Fig.~\ref{fig:discussion}a for tuber amount and tuber state, and in Fig.~\ref{fig:discussion}b for soil moisture and nitrogen.

To compare across conditions on a common scale, we standardize each metric using a reference set and compute z scores. Let $m$ denote an aggregated indicator value and let $\mathcal{R}$ denote the reference set used for standardization. We compute
\begin{equation}
z = \frac{m - \mu}{\sigma} ,
\quad
\mu = \frac{1}{|\mathcal{R}|}\sum_{i\in\mathcal{R}} m_i ,
\quad
\sigma = \sqrt{\frac{1}{|\mathcal{R}| - 1}\sum_{i\in\mathcal{R}} (m_i - \mu)^2} .
\end{equation}

Given standardized samples $\{z_i\}_{i=1}^{N}$ for a condition, we estimate a probability density function using a kernel density estimate
\begin{equation}
\widehat{p}(z) = \frac{1}{N h}\sum_{i=1}^{N} K\!\left(\frac{z - z_i}{h}\right) ,
\end{equation}
where $K(\cdot)$ is a Gaussian kernel and $h$ is the bandwidth.
When a scalar summary of distribution separation is useful, we compute a symmetric divergence between densities such as Jensen Shannon divergence
\begin{equation}
D_{\mathrm{JS}}(p,q)
=
\frac{1}{2}D_{\mathrm{KL}}(p \| m)
+
\frac{1}{2}D_{\mathrm{KL}}(q \| m) ,
\quad
m = \frac{1}{2}(p+q) ,
\end{equation}
with
\begin{equation}
D_{\mathrm{KL}}(p \| q) = \int p(z)\log\!\left(\frac{p(z)}{q(z)}\right)\, dz .
\end{equation}

These perturbations are designed to modulate the electromagnetic loss and scattering environment experienced by the LTE link. Changes in soil moisture primarily alter effective permittivity and conductivity. A convenient representation is the complex relative permittivity
\begin{equation}
\tilde{\epsilon}_r(f) = \epsilon_r'(f) - j \epsilon_r''(f) ,
\quad
\epsilon_r''(f) \approx \frac{\sigma}{2\pi f \epsilon_0} ,
\end{equation}
which links higher conductivity $\sigma$ to larger loss at frequency $f$. Changes in nitrogen concentration are treated as a controlled manipulation of soil ionic content and related conductive loss. The resulting z score densities provide a compact way to summarize how cellular indicators shift under each controlled condition, which we interpret in the Discussion using the representative plots in Fig.~\ref{fig:discussion}.

For perturbation plots, we compare metric distributions using z scores,
\begin{equation}
z = \frac{m - \mu}{\sigma} .
\end{equation}

\subsection*{Evaluation metrics}

For 5~cm-grid localization, we compare the predicted score map $\widehat{y}(g)$ with the binary ground truth $y(g)\in\{0,1\}$ over grid cells $g\in\mathcal{G}$.
We report: (i) cell-wise accuracy (with an optional distance tolerance), and (ii) two continuous map metrics (MSE and SSIM) used in Fig.~2d.

\begin{itemize}
\item \textbf{Accuracy (threshold $\eta$).} We binarize the score map at threshold $\eta$ and compute the fraction of correctly labelled cells:
\begin{equation}
\mathrm{Acc}(\eta)
=
\frac{1}{|\mathcal{G}|}
\sum_{g\in\mathcal{G}}
\mathbb{I}\!\left[
\mathbb{I}\!\left(\widehat{y}(g)\ge \eta\right) = y(g)
\right].
\end{equation}
When reporting accuracy ``within $d$~cm'', we dilate the ground-truth occupancy map by radius
$r=\left\lceil d / 5~\mathrm{cm}\right\rceil$ before computing $\mathrm{Acc}(\eta)$.

\item \textbf{Mean squared error (MSE).} We quantify continuous error between the score map and the binary labels:
\begin{equation}
\mathrm{MSE}
=
\frac{1}{|\mathcal{G}|}
\sum_{g\in\mathcal{G}}
\left(
\widehat{y}(g) - y(g)
\right)^2 .
\end{equation}

\item \textbf{Structural similarity index (SSIM).} We report SSIM to capture structural agreement between the predicted and ground-truth maps~\cite{wang2004ssim}.
(Parameters follow the standard definition used in this work.)
\end{itemize}

Unless otherwise stated, each pot contributes one LTE grid scan for localization evaluation and one CFR sweep per day for 45~days for growth monitoring.

\subsection*{Statistics and reproducibility}
We distinguish biological and technical replicates to align with Nature Communications statistical reporting expectations. In the 45~day greenhouse study, each pot constitutes one biologically independent replicate ($n=4$ pots), while daily CFR sweeps (45 per pot) and within-position LTE samples (100 per grid position per scan) are technical replicates used to reduce measurement noise. For plots reporting mean$\pm$s.d., the caption specifies the corresponding $n$ (e.g., $n=100$ samples per position for LTE indicators). For growth-stage classification, we use leave-one-pot-out cross validation, training on three pots and evaluating on the held-out pot; reported accuracies are averaged across the four held-out folds. For feature--harvest associations, $n=4$ limits inferential strength; we therefore report effect sizes with bootstrap confidence intervals (10{,}000 pot-level resamples) and treat $p$ values as descriptive. No formal multiple-comparison correction is applied because the association analysis is exploratory.

\section*{Data availability}


Source data underlying the main figures and tables are provided with this paper. Raw and processed CFR sweeps, LTE grid scans, and tuber ground-truth annotations are available at \href{https://drive.google.com/file/d/1zOmmNdH-MWiQSmJ8J-Bz-LQz_wd_v_RE/view?usp=sharing}{this repository}. Additional materials are available from the corresponding author upon reasonable request.


\section*{Code availability}

All code used for data acquisition, preprocessing, feature extraction, and LTE metric fusion is available at \url{https://github.com/mli55/SWAMP}. The repository includes scripts and configuration files required to reproduce the main results and figures reported in this study, together with example commands and environment specifications.

\section*{Acknowledgements}
This work was supported by the \textbf{NC State Plant Sciences Initiative}. We thank the NC State Phytotron for greenhouse facilities and technical support.

\section*{Author contributions statement}
All authors conceived the study. M.L. and T.F. designed the experimental protocol. T.F. conducted the greenhouse experiments, collected the RF and LTE datasets, and performed data preprocessing. M.L. carried out the data analysis and drafted the manuscript. All authors discussed the results, revised the manuscript, and approved the final version. W.W. supervised the project.

\section*{Competing interests}
The authors declare no competing interests.

\end{document}